\documentclass[onecolumn,showpacs,showkeys,superscriptaddress]{revtex4}

\usepackage{amsmath}
\usepackage{bbold}
\usepackage{amsfonts}
\usepackage{amssymb}
\usepackage{pbsi}
\usepackage[T1]{fontenc}
\usepackage{xcolor}
\usepackage{graphicx}
\usepackage{subfig}
\usepackage{float}
\usepackage{color}
\usepackage{hyperref}
\usepackage{multirow}
\usepackage{soul}
\hypersetup{colorlinks = true,
	linkcolor = red,
	anchorcolor = blue,
	citecolor = blue,
	filecolor = blue,
	urlcolor = blue}

\def\openone{\leavevmode\hbox{\small1\kern-3.8pt\normalsize1}}
\def\N{\leavevmode\hbox{ Z \kern-8 pt\normalsize{Z}}}
\def\openone{\leavevmode\hbox{\small1\kern-3.8pt\normalsize1}}
\def\openJ{\leavevmode\hbox{J \kern-9.5pt\normalsize J}}
\def\openS{\leavevmode\hbox{ S \kern-9.3pt\normalsize S}}

\newcommand{\bb}{\begin{equation}}
\newcommand{\ee}{\end{equation}}
\newcommand{\eqb}{\begin{eqnarray}}
\newcommand{\eqf}{\end{eqnarray}}

\begin{document}


\title{Laser-Driven Annular Shock Waves as Laboratory Analogues of $w$CDM Cosmologies and Cosmological Gravitational Waves}

\author{Felipe A. Asenjo}
\email{felipe.asenjo@uai.cl}
\affiliation{Facultad de Ingenier\'ia y Ciencias,
Universidad Adolfo Ib\'a\~nez, Santiago 7491169, Chile.}
\author{Felipe Veloso}
\email{fveloso@uc.cl}
\affiliation{Instituto de F\'isica, Pontificia Universidad Cat\'olica de Chile, Santiago 7820436, Chile.}
\author{Julio C. Valenzuela}
\email{jcvalen1@uc.cl}
\affiliation{Instituto de F\'isica, Pontificia Universidad Cat\'olica de Chile, Santiago 7820436, Chile.}

\begin{abstract}
We demonstrate that the experimental evolution of an annular, laser-driven plasma shock wave, expanding over time and undergoing self-interaction gives rise to multiple shock structures that evolve analogously to a multicomponent cosmological universe. Different propagation trajectories along the shock surface correspond to various forms of $w$CDM cosmologies, enabling the study of scenarios ranging from simple radiation- or matter-dominated universes to those including dark energy. We further show that the dynamics of the Mach stems approximately follows a Hubble-like law. Additionally, perturbations in the shock fronts serve as experimental analogues of cosmological gravitational perturbations in a matter-dominated universe. This work opens a new experimental pathway for classically simulating complex cosmological models and gravitational waves at macroscopic scales in the laboratory.
\end{abstract}


\maketitle

\section{Introduction}
\label{intro}

Cosmology is a science that relies on observation. Because of such limitation, the current status on our understanding of the Universe presents several challenges. Some few examples are the Hubble tension problem  \cite{divalentino,rida,Pedrotti}, evolving dark energy \cite{desi},
or the possible dipolar distribution of the dark energy density \cite{clochi}.   Therefore, any possible realization of an analogue experiment on cosmological scenarios may be helpful to open new veins on the comprehension
of the dynamics of the Universe. This work is devoted to show that these analogue cosmological experiments can be performed using annular plasma shock waves.

The physics of shock waves is well-established since the 20th mid-century \cite{zeld,Sedov,Drake}. 
From the countless different physics experiments where shock waves can be produced and used, plasma shock waves have the interesting feature that they can be produced on mesoscopic scales, in relatively inexpensive experimental configurations \cite{veloso}. Most importantly, they are classical in nature and thus can be studied by simple Newtonian physics. This implies that plasma shock waves can be studied in almost any plasma experimental facility in the world. There is where it lies its importance.

On the other hand, recently, several aspects of cosmology have been showed to exist in analogue experimental systems \cite{barcelo,Chatrchyan,Steinhauer,Viermann,Fifer,Eckel,Banik,Hung,rhyno}. They allow to replicate or produce several aspects of the dynamics of different cosmological Universes, thus allowing to study cosmology in a complementary manner to  observations. Those works use the properties of  quantum systems in order to mimic cosmological features on quantum fields. However, and differently, in this work  we show that plasma shock waves can mimic, at classical level, different kinds of cosmologies, and cosmological features. Below, we present a series of results that show that plasma shock waves can be used to analogue study simple Friedman-Lema\^itre-Robertson-Walker (FLRW) cosmological models, diverse $w$CDM cosmological scenarios, and cosmological gravitational waves. The classical nature of the analogue system is what differentiate our proposal from all the previous one. It allows to reproduce analogue dynamics for cosmological scenarios at a macroscopical scale.

The analogy between shock waves and cosmology  is possible by the one-to-one correspondence between the theory for shock wave evolution, and the Einstein equations for an isotropic cosmological Universe. 
The Sedov-Taylor theory for shock waves \cite{zeld,Sedov,Drake} consider that the mass of a $\zeta$-dimensional blast is    $M=\rho\, r^\zeta$, where $\rho$ is the
  $\zeta$-dimensional constant density of the whole expanding mass (for instance, for a spherical blast, $\zeta=3$), and  $r$ is the radius of the blast. As it propagates, the energy $E=M \dot r^2$ of the blast wave
is constant (here $\dot{} \equiv d/dt$ stands for the time derivative). From here (see Supplemental material) we can deduce that the radius of the shock wave follows a dynamics given by the equation $2 {\ddot r}/{r}+\zeta ({\dot r}/{r})^2=0$,  that has as a solution for the radius $r(t)=r_0\,  t^{2/(2+\zeta)}$, where $r_0=\left[{(2+\zeta)^2 E}/{4\rho}\right]^{1/(2+\zeta)}$ is a constant related to the energy. Therefore, the evolution of the shock waves is controlled by its energy and its spatial dimension. 

On the other hand, in a standard FLRW cosmology \cite{weinberg,ryden}, the Einstein equation for an isotropic Universe can be put in a single form as $2{\ddot a}/{a}+(1+3w)({\dot a}/{a})^2=0 $, where  $a(t)$ is the scale factor of the Universe, and $w=p/\epsilon$ is the equation of state of the content of the Universe (composed by a fluid with energy density $\epsilon$ and pressure $p$). The scale factor results to have the form $a(t)=a_0\, t^{2/(3+3w)}$, with constant $a_0$. 

From the above equations and solutions, the analogy between
shock waves and a simple cosmological model is now evident. The radius $r$ of the shock wave plays the analogue role of the scale factor $a$ of the Universe, and the dimension of the shock wave corresponds to the different types of contents of the Universe, by the relation 
\begin{equation}
\zeta=1+3w\, .
\label{analogyzetaw}
\end{equation}For instance,
a planar shock wave ($\zeta=1$) is the analogue to  a  cosmological Universe filled with dust ($w=0$, non-relativistic matter). A cylindrical shock wave ($\zeta=2$) corresponds to a Universe filled with radiation ($w=1/3$, ultra-relativistic matter). Similarly, a spherical shock wave ($\zeta=3$) analogously describe 
a Universe filled with ultra-relativistic form of radiation ($w=2/3$).

By experimentally constructing  shock waves evolving with different dimensions, it is possible to analogously study several cosmological models and their consequences. As the dimension $\zeta$ is not restricted to be an integer, there is a lot of freedom to reproduce complex cosmological scenarios. This is what we present in this work.
In the following we show that the experimental dynamics of an initial annular (cylindrical) plasma shock wave,  self-interacting  as it evolves, modifies its dimension allowing to describe several  aspects of more complicated cosmologies, such as those described by a  $w$CDM model,  corresponding to an isotropic extension of the $\Lambda$CDM model by including a dark energy fluid with varying equation of state. 
It is possible to show that some shock wave features has
fractional negative dimensions $0<\zeta<1$, mimicking the effect of a Universe filled with a content composed by accelerating dark energy, with $w<-1/3$. This takes place in the triple point of the plasma shock wave evolution, which occurs by the interaction of   incident and reflected cylindrical shock fronts, forming a third shock wave, known as a Mach stem \cite{veloso}.

All this implies that the complete evolution of the complex shock wave can be modeled by a phenomenological combination of their different stages, as the wave develops and evolves in time. Therefore, using the previous discussed theory for the evolution of a shock wave for a single dimension,
the complex temporal change of the  shock wave radius is proposed to satisfy the equation $\dot r^2=\sum_i E_i/M_i$, where the index $i$ stands for the different stages of evolution of the wave, composed by shock waves with different energies and masses $E_i$ and $M_i$, respectively. As $M_i=\rho_i r^{\zeta_i}$, we conclude that  a complex shock wave evolution must follow the equation  
\begin{equation}
    \left(\frac{\dot r}{r}\right)^2={\sum_i\frac{\Omega_i}{r^{\zeta_i+2}}}\, ,
    \label{analogelambdacdmshock}
\end{equation}
with the constants for each stage $i$
\begin{equation}
    \Omega_i\equiv \frac{E_i}{\rho_i}=\frac{4 r_{0i}^{2+\zeta_i}}{(2+\zeta_i)^2}\, .
    \label{Omegaistage}
\end{equation}

Eq.~\eqref{analogelambdacdmshock} is the analogue equation to the flat $w$CDM cosmological model $(\dot a/a)^2={(\Omega_b+\Omega_c)/a^3+\Omega_r/a^4+\Omega_\Lambda/a^{3+3 w}}$ \cite{weinberg,yadav}, where $\Omega_j$ are the present-day density parameters for baryons ($j=b$), cold matter ($j=c$), radiation ($j=r$) and dark energy ($j=\Lambda$). For the evolution of the shock wave, described by Eq.~\eqref{analogelambdacdmshock},  different dimensions of propagation are the corresponding analogues to different contents of the Universe, as it evolves. Besides, $\dot r/r$ plays the role of an effective Hubble parameter for the shock wave dynamics.

In the following, we experimentally show that the dynamics of different parts of a laser produced annular plasma shock wave indeed follows Eq.~\eqref{analogelambdacdmshock}, as they develop through different stages.
These different evolving parts define paths, where their expanding radius $r$  can be followed in time. The experimental data for the shock wave expansion  of  the different paths' radius are shown to fit the Sedov-Taylor theory or the
Eq.~\eqref{analogelambdacdmshock},  with a correlation coefficient $R^2 > 0.997$ for all cases.

\begin{figure}
    \centering
    \includegraphics[width=0.9\linewidth]{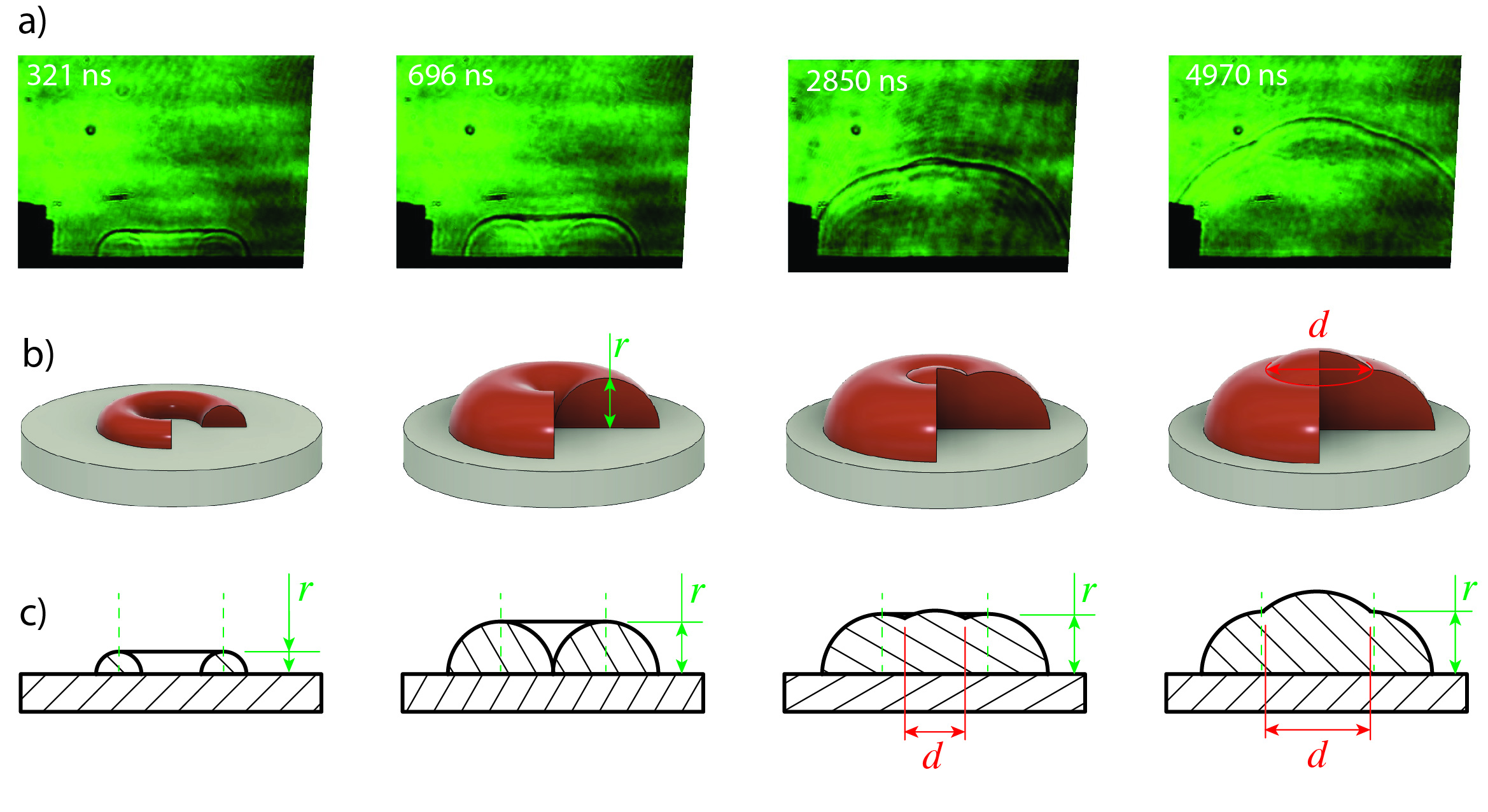}
    \caption{Temporal sequence of the laser induced annular shock wave. (a)  Experimental schlieren images of the shock, for four different representative times. (b) Simulated three-dimensional reconstructions of the shock wave propagation for the same four times. (c) Cross-section view of the shock wave dynamics for the same times. Here $r$ is the radial distance of the annular (cylindrical) shock wave, and $d$ is the distance between the triple points.}
    \label{shock1}
\end{figure}

The laser-produced annular plasma shock wave is generated when an axicon prism is located in the path of a focusing high power laser beam \cite{veloso2006}. In this situation, in the focal plane of the converging lens, a ring-shaped focusing of the laser is achieved in contrast to a focal point as typically occur when no axicon is present. When locating a target in the focal plane in a controlled background atmosphere, the laser produced annular plasma acts as an annular piston to drive cylindrical shock waves. For the experiment analyzed in this paper, we use a Nd:YAG laser ($\lambda=1064$nm, 90mJ, 8ns FWHM) forming a ring-shaped focus of $R_0=2.5$mm on an aluminum target in a 100 torr argon background. As the laser induced annular shock wave propagates, it evolves forming different shock structures, as shown in Fig \ref{shock1}(a).  The images were obtained using a bright-field schlieren imaging system to visualize the shock wave. A frequency-doubled probing laser beam (532 nm, 8 ns) was sent parallel to the target surface and delayed in time to capture different stages of the shock evolution. Refractive index gradients caused by the shock front deflected the probing beam, and when this deflection exceeds the acceptance angle of the schlieren aperture, the light is blocked, resulting in darker regions in the image that reveal the structure and evolution of the shock.

A detailed description of the experiment can be found elsewhere \cite{veloso}, but a brief description is provided here. 
Firstly, the annular shock front follows the expansion expected for an annular piston, i.e., expansion similar to a 'donuth' (see first images in rows a, b and c in Fig.~\ref{shock1}). Later, at the center of the annular configuration, diametrically opposed shock fronts encounter each other at the axis of the configuration (second images in rows a, b and c in Fig.~\ref{shock1}). When this encounter occurs at a certain angle, incident and reflected shock fronts produce a tertiary shock, known as Mach stem, which permits the flow conservation to be fulfilled \cite{ben-dor} (third images in rows a, b and c in Fig.~\ref{shock1}). The intersection of these three shock fronts is known as triple point, which are indicated by the red vertical lines in Fig.~\ref{shock1}. Now, the complete shock front structure is a combination of the original annular plasma-driven shock front combined with the Mach stem (fourth images in rows a, b and c in Fig.~\ref{shock1}). 
At much later times, the combined front propagates at distances much larger than the  initial diameter of the annular configuration and therefore, the propagation  approaches to a spherical shock wave. Hence,
the whole evolution of the laser produced annular plasma transits from an initial cylindrical shock wave to a final state of a spherical shock wave, with a intermediate phase of Mach stem evolution. These changes in the geometry of the complete shock front result in changes in the dimension described by the parameter $\zeta$.

In order to evaluate the laser induced annular shock waves as a classical mechanics analogue experiment for cosmological models, we analyze different paths over the surface of the shock front to describe different cosmological models. First, we will focus in the path described by parallel lines perpendicular to the aluminum target (vertical green dashed lines in Fig.~\ref{shock1}). In the laboratory frame of reference, the distance between these lines remains constant, whereas the  surface of the shock front variates. Secondly, we consider the evolving paths of the triple points. Also, we consider
the distance separating such points (distance $d$ in Fig.~\ref{shock1}), representing the distance between two observers, each standing in a triple point. Finally, we consider the evolution of the Mach stem as an induced perturbation of the overall shock front. We demonstrate that all these  cases have specific analogue cosmological interpretations.


\section{Path of parallel lines}

Let us start our discussion on the usefulness of plasma shock waves as analogue cosmological models, by
initially focusing
in the  path of the parallel lines described in Fig.~\ref{shock1} by vertical green dashed lines. These paths have been chosen arbitrarily in order to consider the simplest measurement for the evolving radius due the different interactions of the shock waves. For this case, we call radius $r$ of the shock wave to the growing distance (along the path) from the base to the surface of the shock wave. This is what is depicted by the vertical green dashed lines in Fig.~\ref{shock1}.
As we will see, the shock wave radius of this path measures the changes on shock wave dimensionality as the blast progress in time.

\begin{figure}
    \centering    \includegraphics[width=0.6\linewidth]{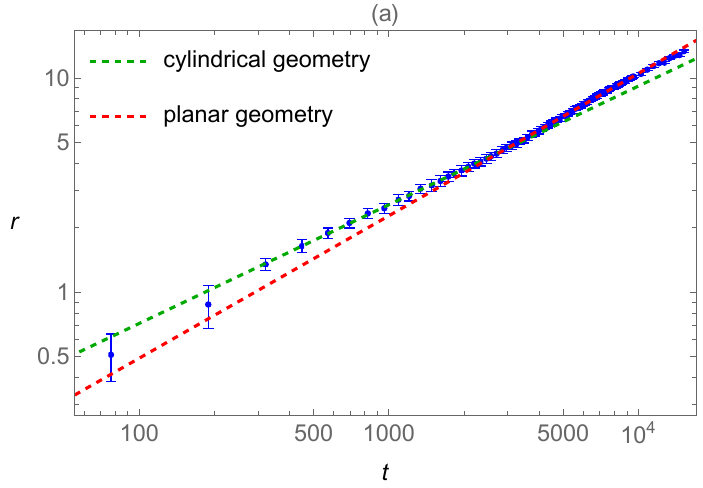}    \includegraphics[width=0.6\linewidth]{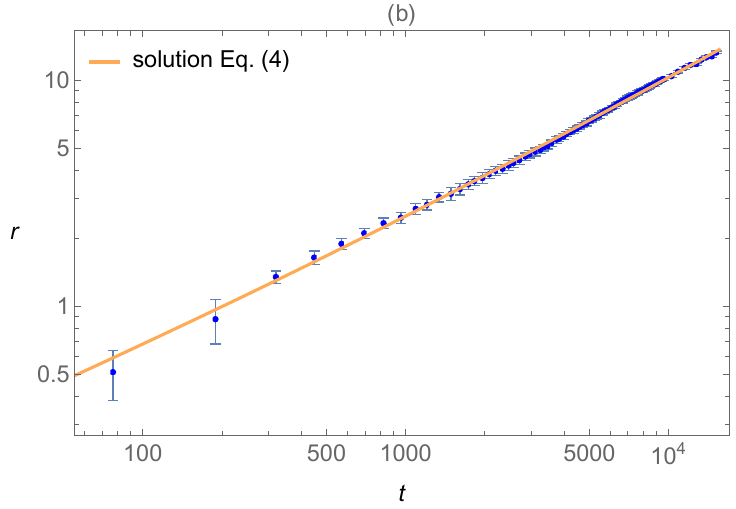}
\includegraphics[width=0.6\linewidth]{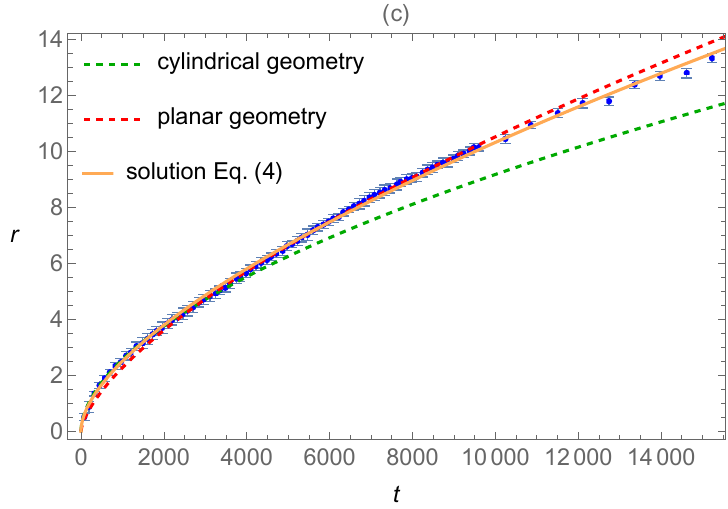}
    \caption{Temporal evolution of the radius of the shock wave in a parallel path configuration (blue circles) with their corresponding error bars. The scale of distance is mm, and time is in ns. (a) Log-log plot for the fitting the data by two different stages in the evolution, a cylindrical  geometry (green dashed line) and a planar geometry (red dashed line). (b) Log-Log plot of the shock wave data  and its fitting by the solution of Eq.~\eqref{ajusteglobalpath1} (orange solid line). (c) Linear scale plot for the shock wave data (blue circles)  with the fit for the two single shock wave stages (cylindrical  geometry in green dashed line and  planar geometry in red dashed line), and the solution of Eq.~\eqref{ajusteglobalpath1} (orange solid line).}
    \label{path1}
\end{figure}

The time dependence of the shock wave radius of this path is shown in Fig.~\ref{path1}(a) in a log-log scale plot, where the evolution of the shock wave should follow the linear equation $\log r=\log r_0+2/(2+\zeta)\log t$, according to the Sedov-Taylor theory. In this plot, and in all the following ones, time is measured in nanoseconds, while the radius is measured mm.
Blue circle dots represent the experimental evolution of the radius of one of the parallel paths. We can see that the dimensionality of the shock wave is not maintained in time, but rather it abruptly changes. This is shown in the numerical fitting in dashed lines.  In this log-log plot, the shock wave initially has a dynamics with  the slope $2/(2+\zeta)=0.554\pm0.004$, and $\log r_0=-2.89\pm0.03$ (green dashed line), 
behaving approximately in a  cylindrical geometry, with dimension $\zeta\approx 1.61$. Abruptly, its dynamics change to a have a slope $2/(2+\zeta)=0.664\pm 0.003$, and $\log r_0=-3.77\pm0.03$ (red dashed line), implying that the shock later develops in a planar geometry, with dimension $\zeta\approx 1.01$.

In terms of the cosmological analogy, using Eq.~\eqref{analogyzetaw}, the shock wave of this path is describing a mimicking evolution of a Universe that pass from an epoch dominated by radiation, with $w\approx 0.2$, to a later stage where non-relativistic matter dominates, with $w\approx 0.003$. This is a description of the current standard model of the cosmological evolution of our Universe, without considering the contribution of a cosmological constant  \cite{weinberg,ryden} .

The approximation of the evolution of the shock wave by two different stages give us the hint that we can  fit the whole dynamics by mimicking a version of a $w$CDM cosmological analogue model described by Eq.~\eqref{analogelambdacdmshock}. We propose that this shock wave evolves from a cylindrical geometry to a planar one. 
In order to do this, we need to estimate the $\Omega$-values, using Eq.~\eqref{Omegaistage} with the values of a single shock wave. For the initial cylindrical shock wave fitting, $\log(r_0)\approx -2.89$, implying that the $\Omega$-value  of this stage is of the order of  $\sim 9.2\times 10^{-6}$. For the following planar stage,  $\log(r_0)\approx -3.77$, and thus its $\Omega$-value is of the order $\sim 5.3\times 10^{-6}$. Thus, the complete fitting of the data for the shock wave radius $r$ of the parallel path is obtained by assuming that its dynamics follows  equation  \eqref{analogelambdacdmshock} for two stages (from cylindrical wave with $\zeta=2$, to a planar wave with $\zeta=1$), given by
\begin{equation}
    \left(\frac{\dot r}{r}\right)^2={\frac{\Omega_{cyl}}{r^4}+\frac{\Omega_{pla}}{r^3}}\, ,
    \label{ajusteglobalpath1}
\end{equation}
where $\Omega_{cyl}$ and $\Omega_{pla}$ are the corresponding $\Omega$-values for the cylindrical and planar stages, respectively. Notice that Eq.~\eqref{ajusteglobalpath1} is the one-to-one analogue to the $w$CMD model for a cosmology evolving from a radiation-dominated Universe to a matter-dominated one, neglecting dark energies. In this case, $\Omega_{cyl}$ is the analogue to the density parameter for the radiation component of Universe, as well as $\Omega_{pla}$ is the analogue to the density parameter for barionic or cold matter components.

In Fig.~\ref{path1}(b) we show the fitting of Eq.~\eqref{ajusteglobalpath1} in log-log scale (in orange solid line). This is achieved by $\Omega_{cyl}\approx 3.6\times 10^{-6}$, and $\Omega_{pla}\approx 4.1\times 10^{-6}$, which are of the same order of magnitude  to the previous estimations using the relations for single shock waves.
Similarly, in Fig.~\ref{path1}(c) we show the data of the parallel path  (blue circles) in a linear scale, with
the fitting given by Eq.~\eqref{ajusteglobalpath1}  (orange solid line) and the two fittings of Figs.~\ref{path1}(a) for single shock wave dynamics. We can see that the complete dynamics is described only by the solution of Eq.~\eqref{ajusteglobalpath1}. 
Thereby,  by fixing the path to parallel lines in an initial cylindrical shock wave, we can analogue model  the simplest standard cosmology of our Universe.

\section{Path of triple points}

The above analysis can be used to examine other possible paths for the radius along the shock wave expansion. This the case for the formation of the almost-cylindrical symmetric triple points of the system.
They are indicated by the intersection of the shock wave surface and the red vertical  lines in the third and fourth images in rows a, b and c in Fig.~\ref{shock1}.

\begin{figure}
    \centering    \includegraphics[width=0.9\linewidth]{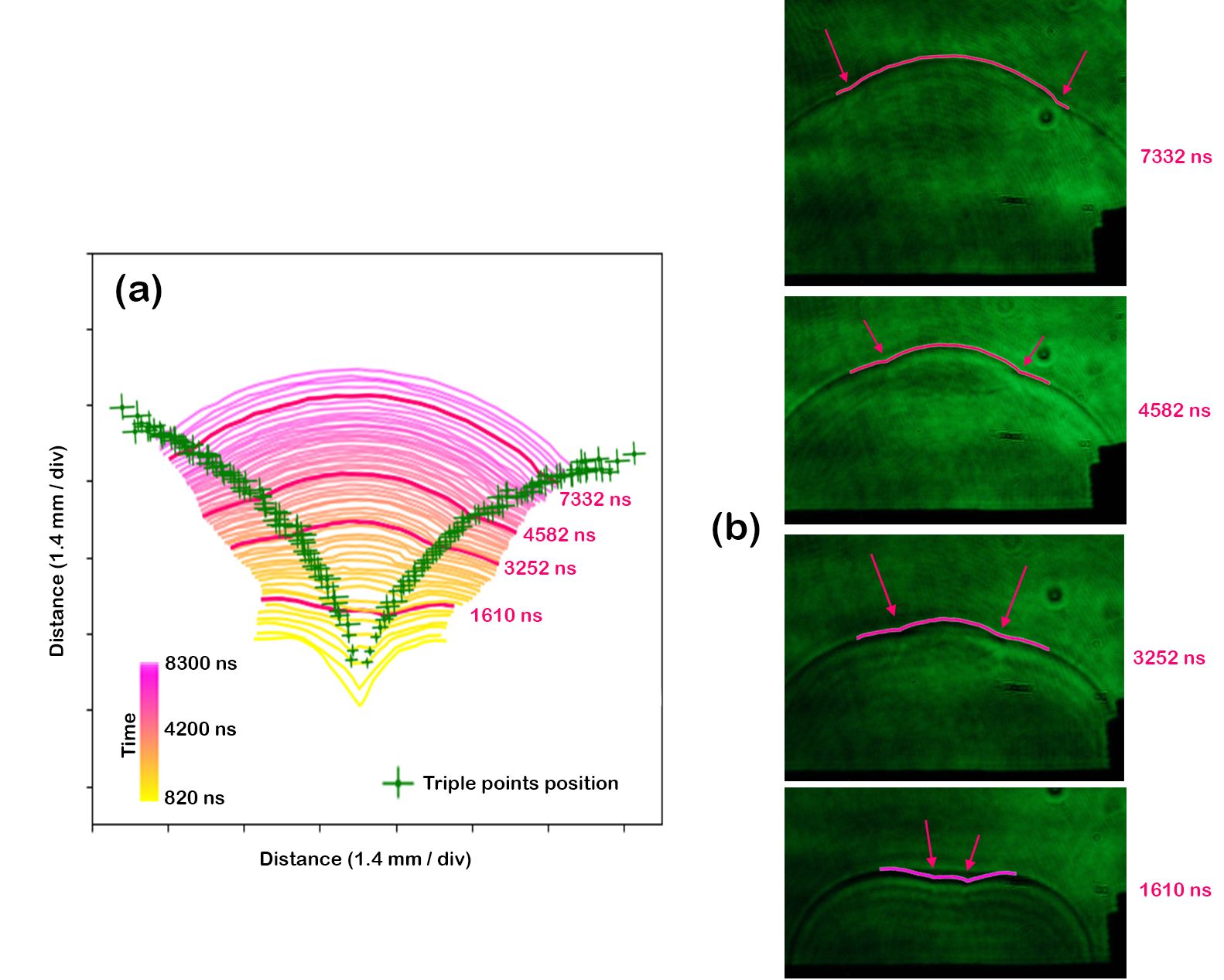}\\
        \caption{(a) Temporal evolution of paths along shock surface (continuous lines) and triple points (green crosses).
        Magenta lines are selected times shown in (b). Arrows indicate triple point positions.}
    \label{stemscolored}
\end{figure}

Those triple points are formed when three shock waves interacts,
an incident one, a reflected one, and the Mach stem shock wave (this last one will be relevant for gravitational waves analogue).  In Fig.~\ref{stemscolored}, the track of the  triple points are shown for the  shock wave profiles, as they evolve in time. In Fig.~\ref{stemscolored}(a), we show successive growing of the shock wave surface (as the time grows). The triple points are marked by the green crosses. We can see how the triple points remains almost parallel for all time, as they separate as the shock wave expands. Some selected times for the shock wave surface are shown in magenta lines, which are also displayed in snapshots for the experimental dynamics of the wave  in Fig.~\ref{stemscolored}(b).

Similarly to the previous section, we can obtain the time evolution of the radius  of both triple points. Here, the radius is defined as the accumulative distance  from the base to the triple point at different times, along the path that the evolving triple points describe, depicted by green crosses  in Fig.~\ref{stemscolored}.

\begin{figure}
    \centering    \includegraphics[width=0.6\linewidth]{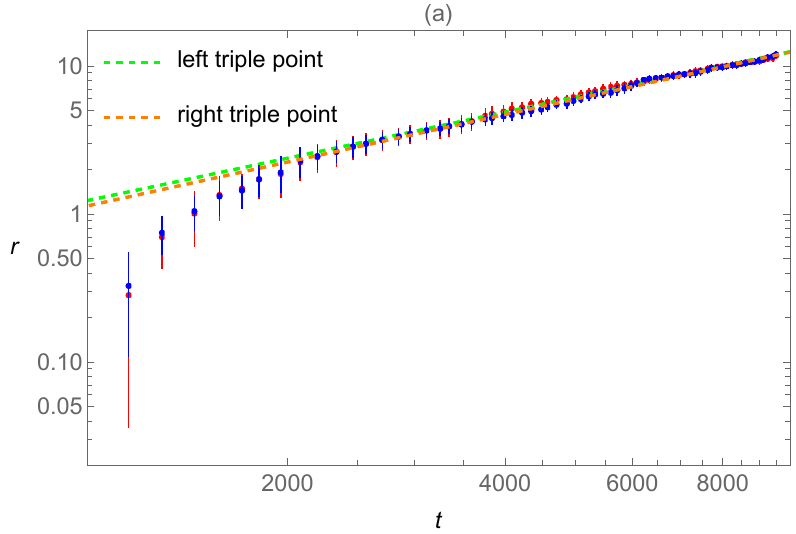}\\
    \includegraphics[width=0.6\linewidth]{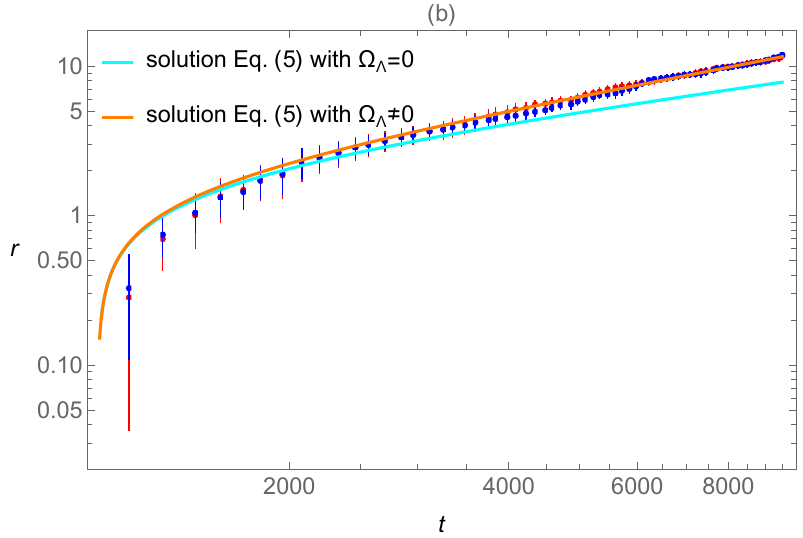}\\
    \includegraphics[width=0.6\linewidth]{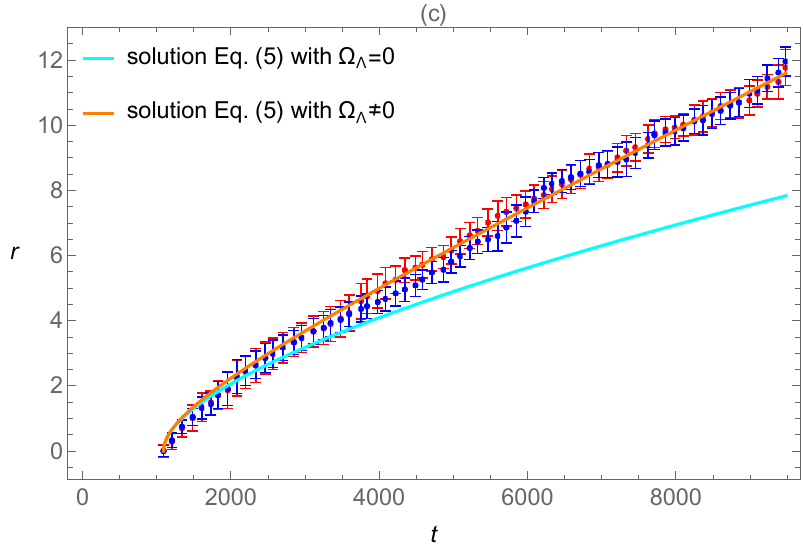}
       \caption{Temporal evolution of the radius of the triple points shock wave;   left (right) triple points in red (blue) circles, 
       with their corresponding error bars. (a) Log-log plot for shock wave fitting for long times. Green (orange) dashed line fitting for left (right) triple point  dynamics. (b) Log-log plot for the best fit (orange line) of Eq.~\eqref{ajusteglobalstems}, with $\Omega_\Lambda\neq 0$, to the triple point data. In cyan line is shown the solution of Eq.~\eqref{ajusteglobalstems} for $\Omega_\Lambda=0$. (c) Same as (b) in linear scale. }
    \label{stems}
\end{figure}

First, we show in Fig.~\ref{stems}(a), the temporal dynamics of this radius (in a log-log scale)  for the left triple point  (in red circles) and the right triple point (in blue circles). They have almost identical dynamics, as the shock wave is symmetric. However, each of them does not fulfill the evolution of a single shock wave.  
In order to display this, in such figure we highlight the behavior of both triple points for long times. Only in that case, the behavior fit the dynamics of a single shock wave discussed in Sec.~\ref{intro}. We can see that left triple point  dynamics (green dashed line fitting) has a power law behavior given by $2/(2+\zeta)\approx 1.03\pm 0.01$, and $\log r_0=-6.99\pm 0.08$.
This implies that, for long times, left triple point behaves as  shock wave evolving with dimension $\zeta\approx-0.06$. On the other hand, right triple point  behaves, for log times, as a single shock wave
with $2/(2+\zeta)\approx 1.07\pm 0.01$ and $\log r_0=-7.32\pm 0.08$, shown by the orange dashed line. Anew, this implies that, for long times, right triple point evolves as  shock wave with dimension $\zeta\approx-0.13$.

These negative dimensions (for long times) represent an accelerated expansion of the triple points, and, by Eq.~\eqref{analogyzetaw}, they are the analogue to the consequence of an accelerating dark energy fluid with equation of state $w=-0.35$ and $w=-0.38$, respectively. Therefore, the evolution of the triple points can be used to describe an analogue system where accelerating dark energy takes place in long times.
In order to fully determine the dynamics of both triple points, and similarly to 
the previous section, we propose that the complete dynamics of each of them can be described by an analogue equation  to the  $w$CDM model, with the structure of   Eq.~\eqref{analogelambdacdmshock}. 
For this case, the equation should acquire the form
\begin{equation}
    \left(\frac{\dot r}{r}\right)^2={\frac{\Omega_{cyl}}{r^4}+\frac{\Omega_{pla}}{r^3}+\frac{\Omega_\Lambda}{r^{2+\zeta_\Lambda}}}\, ,
    \label{ajusteglobalstems}
\end{equation}
which consider  the whole evolution of the triple points, passing from a cylindrical geometry (characterized by $\Omega_{cyl}$) to a planar geometry (characterized by $\Omega_{pla}$), to later been dominated by the  accelerating component with dimension $\zeta_\Lambda$ (characterized by $\Omega_\Lambda$).

The best fit for the solution of Eq.~\eqref{ajusteglobalstems} is shown in an orange solid line in Fig.~\ref{stems}(b) on a log-log scale, and in Fig.~\ref{stems}(c) on a linear scale. This fit occurs for $\Omega_{cyl}=2.46\times 10^{-6}$ and $\Omega_{pla}=2.33\times 10^{-6}$, which are of the same order than the same parameters calculated in previous section. Also, $\zeta_\Lambda=-0.105$ and $\Omega_\Lambda=8.93\times 10^{-7}$, implying that the contribution of the accelerating part behaves with negative dimension (analogously to an accelerating dark energy contribution with $w=-0.37$). This only   dominates for large times, as $\Omega_\Lambda$ has a smaller magnitude than $\Omega_c$ and $\Omega_p$. Besides, this value of $\Omega_\Lambda$ is in agreement with the one calculated using Eq.~\eqref{Omegaistage} for a single shock wave theory (for instance, for the right triple point, we have that $\log(r_0)\approx -7.32$ implying an associated parameter $\Omega_\Lambda\approx 10^{-6}$).
For comparison, in Figs.~\ref{stems}(b) and (c), we have plotted in solid cyan line the solution of Eq.~\eqref{ajusteglobalstems} for $\Omega_\Lambda=0$, showing that the contribution of the accelerating (dark energy analogue) part is the relevant one to describe the whole evolution of the triple points. 

Thereby, the whole dynamics of the triple points is the corresponding analogue to a $w$CDM cosmological model.

\subsection{Hubble law analogue for large times}

\begin{figure}
    \centering    \includegraphics[width=0.6\linewidth]{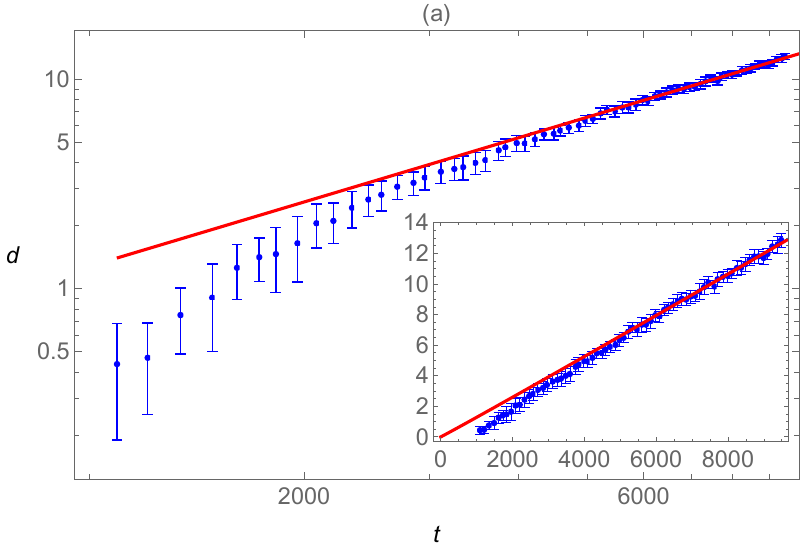}\\
    \includegraphics[width=0.6\linewidth]{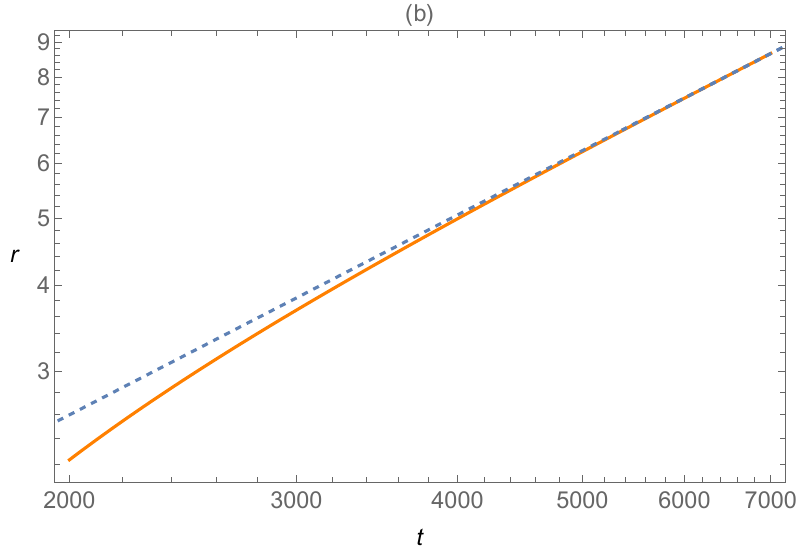}
       \caption{
       (a) In logarithm scale is shown the distance evolution between the triple points (blue dots) compared to its large distance evolution fitting $d \propto  t^{1.02}$ (red line). In the inset plot is shown the fit in linear scale. (b) Best fit solution of Eq.~\eqref{ajusteglobalstems} for the triple point evolution (in orange solid line), altogether with its analytic form (in 
dashed blue line) of its asymptotic  behavior for large times, $r\propto t^{0.9619}$.}
    \label{distanceslarge}
\end{figure}

The evolution of the triple points have also encoded an analogue for the Hubble law.
From the evolution of the positions of these points, we can obtain the time dependent distance $d$ between them, shown in Fig.~\ref{shock1}. This distance is defined as the measurement of the straight line connecting each triple point at given times. 
In Fig.~\ref{distanceslarge}(a)  are shown the data for this distance in blue dots, in a logarithm scale, and in a linear scale (inset plot). At large times, the temporal evolution of distance behaves in the form $d\approx (113\pm 3)\times 10^{-5}\, t^{1.02\pm 0.02}$, which is represented by the red solid line in both plots. This implies that, at large times, we find that $\dot d/d=(1.02\pm 0.02)/t$.

On the other hand, in Fig.~\ref{distanceslarge}(b), we reproduce the complete previous best fit solution of Eq.~\eqref{ajusteglobalstems} for the triple point evolution (in orange solid line), but now showing (in 
dashed blue line) the analytic form of its asymptotic  behavior for large times, that goes as $r\approx(173336\pm 6)\times 10^{-7} t^{0.9619\pm 0.0002}$. This implies that, at large times, the best fitting of Eq.~\eqref{ajusteglobalstems} satisfies
$\dot r/r\approx (0.9619\pm 0.0002)/t$.

The two above results allow us to calculate the relation between the temporal evolution of the distance $d$ and dynamical radius $r$ of the expansion of the triple points. At large times, this results to be
\begin{equation}
  v=\dot d= \left(1.06\pm0.02\right)\left(\frac{\dot r}{r}\right)d\, .
\end{equation}
This is an approximated analogue Hubble law for the recession velocity $v$ between the two triple points. Here, the Hubble parameter $\dot r/r$ is determined by the expansion rate of the triple point radius.

\section{Perturbations in shock wave dynamics as cosmological gravitational wave analogues}

As several paths of the shock wave display different analogue evolutions of cosmological models, we can infer that perturbations to the shock propagation will be the analogue equivalent to perturbations to the cosmological spacetime, i.e., equivalent to cosmological gravitational waves. 

In general, the perturbation of the evolution of a single shock wave radius can be put in the form $r=r_{{u}}(1+\epsilon h)$, where $h$ describe the first order perturbation to the shock wave,  $r_u$ is the unperturbed zeroth-order single shock wave solution described in Sec.~\ref{intro}, and $\epsilon\lll 1$ is  an arbitrary small perturbation parameter. Using the  Sedov-Taylor theory for a single shock wave, we find the equation for the first order perturbation,
$\ddot h+({2}/{t})\dot h=0$,
which is independent of the dimension $\zeta$ (see Suplemental material). The solution for the perturbation is then
\begin{equation}
    h=\frac{h_0}{t}\, ,
    \label{firstorderperturbation}
\end{equation}
independent of the geometry of the propagation of the zeroth-order shock wave ($h_0$ is a constant).  Thus, any shock wave first order perturbation decays in time in an inverse way.

Interestingly, and on the other hand, the tensor (radiative) modes for cosmological gravitational waves follow the equation $\ddot D_{ij}+3({\dot a}/{a})\dot D_{ij}=0$ \cite{weinberg}, when its  spatial variations are negligible, and the
 cosmological anisotropic inertia tensor is not considered. Thus, for the scale factor $a(t)=a_0\, t^{2/(3+3w)}$, the general mode solutions for cosmological gravitational waves are $D_{ij}\propto t^{(w-1)/(w+1)}$.
From here, the analogy  between the perturbations \eqref{firstorderperturbation} on the radius of the shock wave (in any dimension) and cosmological gravitational waves is evident. This analogy is only exact for $w=0$ (non-relativistic matter dominated Universe). 

\begin{figure}
    \centering
    \includegraphics[width=0.8\linewidth]{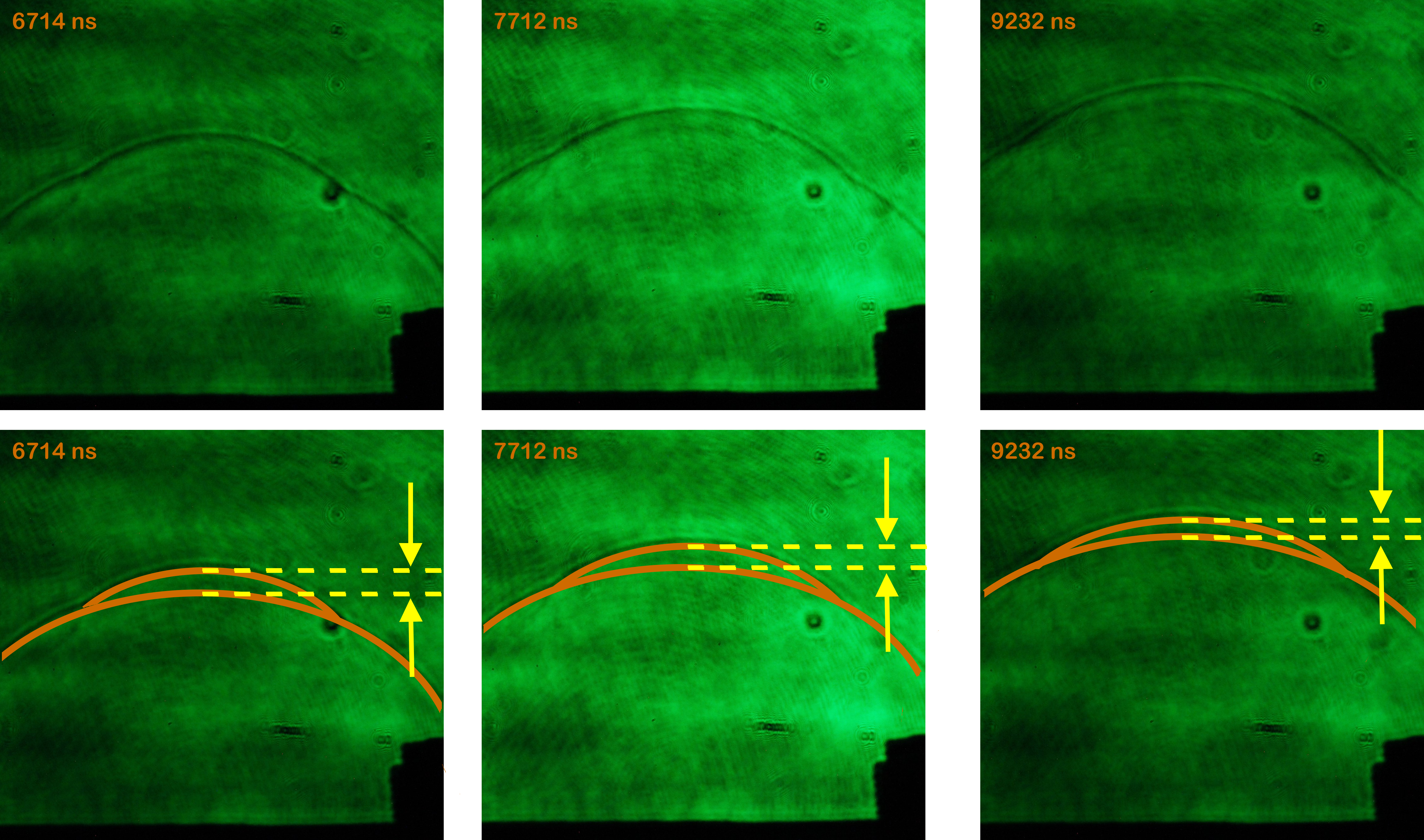}
    \caption{Mach stem evolution as a perturbation of a main shock front. (Top) Raw images. (Bottom) Scheme of main shock and perturbation. Perturbation amplitude is shown with yellow lines}
    \label{shockperturbation}
\end{figure}

Thus, any perturbation  in a single shock wave will be an analogue model to
a cosmological gravitational wave in a Universe similar to the present-day one. In this section, we show that, indeed, the perturbations of the shock wave follow the first-order solution 
\eqref{firstorderperturbation}. This perturbation emerges from the final evolution of the Mach stem of the shock wave, shown in  fourth images of row (a), (b) and (c) of Fig.~\ref{shock1}. The evolution of this perturbation is also shown in
Fig.~\ref{shockperturbation}.   As the triple points shows acceleration (with respect to the rest of the shock wave surface, see Fig.~\ref{stemscolored}) the Mach stem eventually surpasses the  almost-spherical geometry of the total shock wave, creating a ``bump" that has its own temporal evolution. This is shown in the top row of  Fig.~\ref{shockperturbation}. The definition for the amplitude of the radius size  of such bump is depicted in the bottom row of  Fig.~\ref{shockperturbation}, which is the distance between the maximum height of the bump and  the radius of the  almost-spherical   shock wave. As we can see, the bump radius can be considered as a perturbation of the whole shock wave radius dynamics. First it grows, to eventually decays to vanish. Below we show that it is its initial decaying behavior which fulfills the the inverse temporal law described by Eq.~\eqref{firstorderperturbation}.

\begin{figure}
    \centering    
    \includegraphics[width=0.65\linewidth]{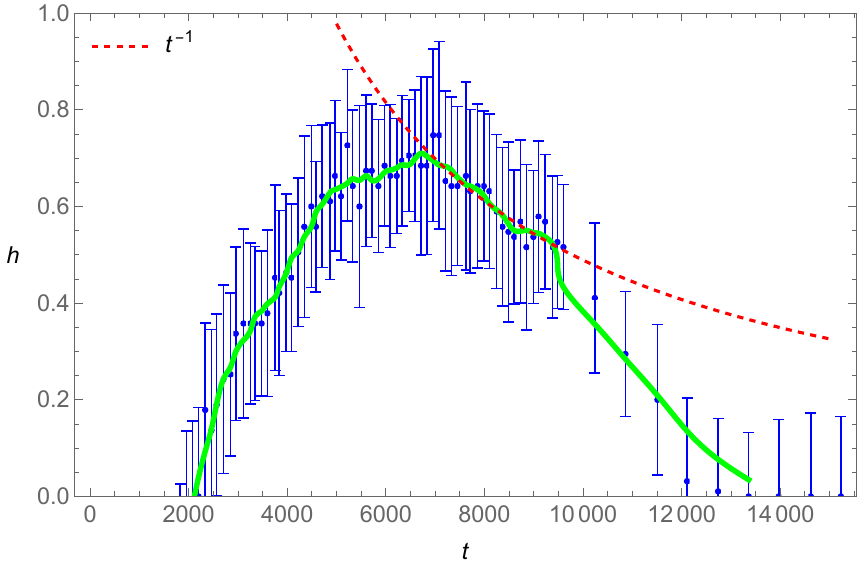}\\
       \caption{Temporal evolution (blue points) of the distance radius of the perturbation amplitude shown in Fig.~\ref{shockperturbation}. In green solid line is shown the moving average of the data. In dashed red line is shown the fit for the decaying, $h\propto 1/t$.}
    \label{gravwaves}
\end{figure}

The data for the dynamics of the temporal evolution of the radius of the perturbation is displayed by blue points (with their respective error bars) in the plot of Fig.~\ref{gravwaves}, in a linear scale. 
The solid green line represents the moving average of the data. The dynamics of the perturbation start from the formation and growing of the stem, to later  decrease at approximated time $t\approx 6900$ ns. It is from that time, that the decaying behavior 
of the perturbation is the analogue to the cosmological gravitational wave perturbation. This is shown by the dashed red line,  corresponding to the fitting $h=h_0 t^\lambda$, where $\lambda\approx=-1.00\pm 0.03$, and $h_0=4900\pm 170$, in agreement with the solution \eqref{firstorderperturbation}. In this way, the perturbation of the shock waves start decaying (at first order) as 
predicted by Eq.~\eqref{firstorderperturbation}. Therefore, these perturbations are able to display an analogue behavior to cosmological gravitational waves in a matter-dominated Universe.

For later times, from around $t\approx 9500$ ns, the decaying of the stem shock
wave follows a second-order perturbation dynamics (see Suplemental material).

\section{Final comments}

It is remarkable that the evolution of several parts of a complex shock wave can  classically mimic the temporal evolution of different cosmological Universes and different cosmological features.
This is basically achieved by the one-to-one correspondence between the Sedov-Taylor theory and the isotropic FLRW cosmology. All these analogue agreements are  summarized in Table \ref{tableresumen}.

In principle,  different particular $w$CDM cosmological models can be analogously crafted by some more complex expanding plasma shock waves. For instance, starting from different geometries, or allowing the interaction of several evolving
Mach stems. The same can be said for the perturbations of the shock waves, which can be engineered  to model cosmological gravitational waves for arbitrary geometries.
This is a subject of study right now.

The most interesting inference that can be extracted of these experiments is that the accelerating nature of part of the plasma shock wave (seen in the evolution of the triple points of the Mach stem) is not  fundamental in nature, but a consequence of nonlinear interactions between different parts of the total shock wave. Although its dynamics can be described by an $w$CDM analogue equation, the analogue dark energy contribution emerges simply by the interaction of cylindrical waves, changing its dimensionality in the process, in order to preserve the flow of the wave dynamics \cite{ben-dor}. This is an idea that can have some usefulness on  current studies of the Universe.

We finally stress one of the main contribution of our findings. This is that we can create analogue cosmological settings at a macroscopic classical level.
Thus, by using plasma shock waves, the current results can be replicated, or extended, in almost every simple plasma laboratory in a straightforward manner.

\begin{table}
\caption{Summary of different parameters that allow the analogue correspondence between plasma shock waves and cosmology.}
\label{tableresumen}
\begin{tabular}{|p{4cm}|p{4cm}||p{5cm}|p{4cm}|}
\hline
\multicolumn{2}{|c||}{Laboratory experiment} & \multicolumn{2}{|c|}{Cosmological analogue}\\
\hline \hline
Measured distance & Parameters & Analogue & Parameters \\ \hline 
Parallel lines & $\log r_{0,1} = -2.89 $ & \multirow{2}{5cm}{Evolution from radiation- to matter- dominated Universe} &  $\Omega_{cyl} = 3.6\times 10^{-6}$\\
 & $\zeta_1 = 1.61$ & & $\Omega_{pla}=4.1\times 10^{-6}$\\
 & $\log r_{0,2} = -3.77 $ & & $\Omega_\Lambda= 0$ \\
 & $\zeta_2 = 1.01$ & & \\

 \hline
Triple points trajectories & $\log r_{0L} = -6.99 $ & \multirow{2}{5cm}{Accelerating dark energy taking place at long times} &  $\Omega_{cyl} = 2.46\times 10^{-6}$\\
 & $\zeta_L = -0.06$ & & $\Omega_{pla}=2.33\times 10^{-6}$\\ 
 & $\log r_{0R} = -7.32 $ & & $\Omega_\Lambda=8.93\times 10^{-7}$\\
 & $\zeta_R = -0.13$ & & \\

\hline
Path connecting triple points ($d$) & $ d \propto t^{1.02}$ & \multirow{2}{5cm}{Hubble law for isotropic Universe} & $\dot d =1.06 (\dot r/r)d$\\
 &$r\propto t^{0.96}$ & & \\ 
 & & & \\
\hline
Shock perturbation by Mach stem ($h$) & $\ddot h+(2/t)\dot h=0$ & \multirow{3}{5cm}{Gravitational wave in a non-relativistic matter dominated Universe} & $h=h_0t^\lambda$  \\
 & & & $\lambda = -1.00 \pm 0.03$\\ 
 & & &  \\
\hline
\end{tabular}
\end{table}

\begin{acknowledgements}
FAA thanks to FONDECYT grant No. 1230094. FV thanks to FONDECYT grant No. 1231286 that supported this work. JCV thanks to FONDECYT grant No. 1220533 that supported this work.
 \end{acknowledgements}


\begin{thebibliography}{}

\bibitem{divalentino} E. Di Valentino
{\it et al.},
Class. Quantum Grav. {\bf 38}, 153001 (2021).

\bibitem{rida} A. R. Khalife {\it et al.}, JCAP04, 059 (2024). 
\bibitem{Pedrotti} D. Pedrotti {\it et al.},    Phys. Rev. D {\bf 111}, 023506 (2025).

\bibitem{desi} DESI Collaboration et al. arXiv:2503.14739v2 (2025);  arXiv:2503.14738v2 (2025).

\bibitem{clochi} A. Clocchiatti, \'O. Rodr\'iguez, A. \'Ordenes Morales and B. Cuevas-Tapia, ApJ {\bf 971}, 19 (2024). 

\bibitem{Sedov} L. I. Sedov, {\it Similarity and Dimensional Methods in Mechanics} 
(Academic Press, New York, 1959).

\bibitem{zeld}
Y. B. Zel'dovich and Y. P. Raizer, {\it Physics of Shock Waves and
High Temperature Hydrodynamic Phenomena} (Academic Press,
New York, 1966).



\bibitem{Drake} R. P. Drake, {\it High-Energy-Density Physics: Fundamentals, Inertial
Fusion and Experimental Astrophysics} (Springer-Verlag,
Berlin, Heidelberg, 2006).


\bibitem{veloso} F. Veloso, V. Rosales, M. Favre and J. Valenzuela, Phys. Rev. E {\bf 110}, 065210 (2024).


\bibitem{barcelo} C. Barcel\'o, S. Liberati and M. Visser,  Int. J. Mod.  Phys. D {\bf 12}, 1641 (2003). 

\bibitem{Chatrchyan} A. Chatrchyan {\it et al.}, Phys. Rev. A {\bf 104}, 023302 (2021).

\bibitem{Steinhauer} J. Steinhauer {\it et al.}, Nature Comm. {\bf 13}, 2890 (2022).

\bibitem{Viermann} C. Viermann, Nature {\bf 611}, 260 (2022).

\bibitem{Fifer} Z. Fifer {\it et al.}, Phys. Rev. E {\bf 99}, 031101(R) (2019).

\bibitem{Eckel} S. Eckel {\it et al.}, Phys. Rev. X {\bf 8}, 021021 (2018).
\bibitem{Banik} S. Banik {\it et al.}, Phys. Rev. Lett. {\bf 128}, 090401 (2022).

\bibitem{Hung} C.-L. Hung {\it et al.}, Science {\bf 341}, 1213 (2013).

\bibitem{rhyno} B. Rhyno, I. Velkovsky, P. Adshead, B. Gadway and S. Vishveshwara, Phys. Rev. Research {\bf 7}, L022004  (2025).

\bibitem{weinberg} S. Weinberg, {\it Cosmology} (Oxford University Press, 2008).

\bibitem{ryden} B. Ryden, {\it Introduction to Cosmology} (Addison Wesley, 2003).

\bibitem{yadav} V. Yadav, S. Kumar Yadav and Rajpal,
Physics of the Dark Universe {\bf 46},
101626 (2024).



\bibitem{veloso2006}F. Veloso, H. Chuaqui, R. Aliaga-Rossel, M. Favre, I.H. Mitchell, E. Wyndham, Rev. Sci. Instrum. {\bf 77}, 063506 (2006)

\bibitem{ben-dor} G. Ben-Dor, {\it Shock Wave reflection phenomena, 2nd Ed} (Springer, Berlin, 2007) 

\end{thebibliography}
\end{document}